\documentclass[11pt]{article}
\usepackage{amsmath}
\usepackage{amssymb}
\usepackage[dvips]{graphicx}
\begin{document}
\title{Exact Relativistic Static Charged Dust Disks and Non-axisymmetric Structures}
\author{D. Vogt\thanks{e-mail: danielvt@ifi.unicamp.br}\\
Instituto de F\'{\i}sica Gleb Wataghin, Universidade Estadual de Campinas\\
13083-970 Campinas, S.\ P., Brazil
\and
P. S. Letelier\thanks{e-mail: letelier@ime.unicamp.br}\\
Departamento de Matem\'{a}tica Aplicada-IMECC, Universidade Estadual\\
de Campinas 13083-970 Campinas, S.\ P., Brazil}
\maketitle
\begin{abstract}
The well-known ``displace, cut and reflect'' method used to generate 
disks from given solutions of Einstein field equations is applied to the superposition of two
extreme Reissner-Nordstr\"{o}m black holes to construct disks made of charged dust and also
non-axisymmetric planar distributions of charged dust on the $z=0$ plane. They are symmetric with respect to two
or one coordinate axes, depending whether the black holes have equal or unequal masses, respectively.
For these non-axisymmetric distributions of matter we also study the effective potential for geodesic motion of neutral test particles.

PACS: 04.20.Jb, 04.40.-b, 04.40.Nr
\end{abstract}

\section{Introduction}

Solutions of Einstein's field equations representing disk-like configurations of matter are of great 
astrophysical interest, since they can be used as models of galaxies or accretion disks.  
Disk solutions can be static or stationary and with or 
without radial pressure. Solutions for static thin disks without radial pressure were first 
studied by Bonnor and Sackfield \cite{Bonnor1}, and Morgan and Morgan \cite{Morgan1}, and with 
radial pressure by Morgan and Morgan \cite{Morgan2}. Other classes of static thin disk solutions
have been obtained \cite{Lynden1}--\cite{Bicak2}, while stationary thin disks were studied in \cite{Bicak3}--\cite{Gonzalez1}. An 
exact solution to the problem of a rigidly rotating disk of dust in terms of ultraelliptic functions was reported in \cite{Neugebauer}.
Also thin disks with radial tension \cite{Gonzalez2}, magnetic fields \cite{Letelier1} and both electric 
and magnetic fields \cite{Katz1} have been studied. The non-linear
superposition of a disk and a black hole was first considered by Lemos and Letelier \cite{Lemos2}.
Models of thin disks and thin disks with halos made of perfect fluids were considered in \cite{Vogt}.
The generalization of the ``displace, cut and reflect'' method (Sec.\ \ref{sec_form}) of constructing thick
static disks was considered by Gonz\'{a}lez and Letelier \cite{Gonzalez3}.

An interesting class of solutions of the Einstein-Maxwell field equations are the conformastatic
space-times with charged dust, in which the charge density is equal to the mass density. Therefore, the
matter is in equilibrium because the mutual gravitational attractions are balanced by the electrical repulsions. 
This kind of matter has been called by some authors ``electrically counterpoised dust'' (ECD).
Distributions of ECD in equilibrium in both classical and relativistic theories were studied by Bonnor \cite{Bonnor2}. Models
of oblate and prolate spheroids made of ECD were considered, respectively, in \cite{Bonnor3} and \cite{Bonnor4}. 
G\"{u}rses \cite{Guerses} and Varela \cite{Varela} studied static spheres of ECD. Disk sources for conformastationary
metrics (the stationary version of conformastatic metrics) were considered in \cite{Katz1}. Although one may intuitively expect
that astrophysical objects do not have a net charge, there exists the possibility that electrons escape from
a compact star, leaving behind a positively-charged one (see, for example, \cite{Rosseland,Bally}).

In this paper we apply the well known ``displace, cut and reflect'' method to the superposition of two extreme
Reissner-Nordstr\"{o}m black holes aligned on the $z$ axis to generate static disks of ECD on the $z=0$ plane. Next we
repeat the same procedure to Reissner-Nordstr\"{o}m black holes aligned on the $y$ axis and generate non-axisymmetric
distributions of ECD on the plane $z=0$. We briefly study the effective potential of geodesic motion of neutral test 
particles on these structures. 

The paper is divided as follows. Sec.\ \ref{sec_form} discusses the Einstein-Maxwell equations, the
``displace, cut and reflect'' method and the particular class of conformastat metrics. In Sec.\ \ref{sec_aligned1} 
disks of charged dust are constructed using the superposition of two aligned Reissner-Nordstr\"{o}m black holes and 
in Sec.\ \ref{sec_aligned2} we obtain non-axisymetric distributions of charged dust. We also present some 
analysis of the geodesic motion on these matter distributions. Finally, in Sec.\ \ref{sec_disc}, we summarize our results.

\section{Einstein-Maxwell Equations, Disks and Conformastatic Space-Times} \label{sec_form}

We consider a static space-time with coordinates $(t,x,y,z)$ and a line element of the form
\begin{equation} \label{eq_metric1}
\mathrm{d}s^2=e^{\nu(x,y,z)}\mathrm{d}t^2-e^{\lambda(x,y,z)}\left(
\mathrm{d}x^2+\mathrm{d}y^2+\mathrm{d}z^2 \right) \mbox{.}
\end{equation}

The Einstein-Maxwell system of equations is given by
\begin{align}
G_{\mu \nu} &=8\pi T_{\mu \nu} \mbox{,}  \\
T_{\mu \nu} &=\frac{1}{4\pi}\left( F_{\mu}^{\;\; \sigma}F_{ \sigma\nu }+
\frac{1}{4}g_{\mu\nu}F_{\rho\sigma}F^{\rho\sigma}\right) \mbox{,} \label{eq_elm_ten}\\
F^{\mu\nu}{}_{;\mu} &=0 \mbox{,}\\
F_{\mu\nu}&=A_{\nu,\mu}-A_{\mu,\nu} \mbox{,}
\end{align}
where all symbols have their usual meaning. We use geometric units $G=c=1$.

The method used to generate the metric of the disk and its 
material content is the well known ``displace, cut and reflect'' method
that was first used by Kuzmin \cite{Kuzmin} and Toomre \cite{Toomre}
to construct Newtonian models of disks, and later extended to general 
relativity (see, for example \cite{Bicak1,Gonzalez1}).

The material and electric content of the disk will be described by
functions that are distributions with support on the disk. The method can be divided
into the following steps:
First, in a space wherein we have a compact source of gravitational field,
 we choose a surface (in our case, the plane $z=0$) that divides the  space into
two pieces:  one with no singularities or sources and the other with the
 sources. Then we
disregard the part of the space with singularities and use the surface to make an inversion of the
nonsingular part of the space. This results in a space with a
singularity that is a delta function with support on $z=0$.
This procedure is mathematically equivalent to making the transformation
$z \rightarrow |z|+a$, with $a$ constant. In the Einstein tensor we
have first and second derivatives of $z$.  Remembering that $\partial_z |z|=2
\vartheta(z)-1$ and $\partial_{zz} |z|=2\delta(z)$, where
$\vartheta(z)$ and $\delta(z)$ are, respectively, the Heaviside
function and the Dirac distribution, the Einstein-Maxwell equations give us
\begin{align}
G_{\mu \nu} &=8\pi (T_{\mu \nu}^{\mathrm{elm.}}+Q_{\mu\nu}\delta(z)) \mbox{,} \\
F^{\mu\nu}{}_{;\mu} &=4 \pi J^{\nu}\delta(z) \mbox{,} \label{eq_max_disk}
\end{align}
where $T_{\mu \nu}^{\mathrm{elm.}}$ is the electromagnetic tensor Eq.\ (\ref{eq_elm_ten}),
$Q_{\mu\nu}$ is the energy-momentum tensor on the plane $z=0$ and $J^{\nu}$ is 
the current density on the plane $z=0$. 

For the metric Eq.\ (\ref{eq_metric1}), the non-zero components of $Q_{\mu\nu}$ are
\begin{align}
Q^t_t &=\frac{1}{8 \pi}g^{zz}b^x_x \mbox{,} \label{eq_qtt}\\ 
Q^x_x &= Q^y_y =\frac{1}{16 \pi} g^{zz}(b^t_t+b^x_x) \mbox{,} \label{eq_qrr}
\end{align}
where $b_{\mu\nu}$ denotes the jump of the first derivatives of the metric
tensor on the plane $z=0$,
\begin{equation}
b_{\mu\nu}=g_{\mu\nu,z}|_{z=0^+}-g_{\mu\nu,z}|_{z=0^-} \mbox{,}
\end{equation}
and the other quantities are evaluated at $z=0^+$. The electromagnetic potential
for an electric field is
\begin{equation} \label{eq_elec_pot}
A_{\mu}=(\phi,0,0,0) \mbox{.}
\end{equation}
Using Eq.\ (\ref{eq_elec_pot}) and Eq.\ (\ref{eq_max_disk}), the only non-zero
component of the current density on the plane $z=0$ is
\begin{equation} \label{eq_curr_disk}
J^t=\frac{1}{4\pi}g^{zz}g^{tt}a_t \mbox{,}
\end{equation}
where $a_{\mu}$ denotes the jump of the first derivatives of the electromagnetic 
potential on the plane $z=0$,
\begin{equation}
a_{\mu}=A_{\mu,z}|_{z=0^+}-A_{\mu,z}|_{z=0^-} \mbox{,}
\end{equation}
and the other quantities are evaluated at $z=0^+$. The ``physical measure'' of
length in the direction $\partial_z$ for metric (\ref{eq_metric1}) is $\sqrt{-g_{zz}}$,
then the invariant distribution is $\delta(z)/\sqrt{-g_{zz}}$. Thus the ``true'' surface
energy density $\sigma$ and pressures or tensions $P$
are:
\begin{equation} \label{eq_disk_surf}
\sigma=\sqrt{-g_{zz}}Q^t_t \text{,} \quad
P=-\sqrt{-g_{zz}}Q^x_x=-\sqrt{-g_{zz}}Q^y_y \mbox{.}
\end{equation}
Since $J^{\mu}=\hat \rho U^{\mu}$, where $U^{\mu}=\delta^{\mu}_t/\sqrt{g_{tt}}$,
the ``true'' surface charge density $\rho$ is 
\begin{equation} \label{eq_disk_ch}
\rho=\sqrt{-g_{zz}g_{tt}} J^t \mbox{.}
\end{equation}

Let us now specialize metric Eq.\ (\ref{eq_metric1}) to a conformastat form by
considering 
\begin{equation} \label{eq_choice}
e^{\nu(x,y,z)}=V^{-2}(x,y,z) \quad \text{and} \quad e^{\lambda(x,y,z)}=V^2(x,y,z) \mbox{.}
\end{equation} 
With this choice, in the absence of matter, the Einstein-Maxwell equations are
satisfied provided (a) V satisfies Laplace equation and (b) the relation between V and the electric
potential $\phi(x,y,z)$ is of the form (see, for example, \cite{Synge} and the Appendix of \cite{Katz2}
for a detailed deduction)
\begin{equation} \label{eq_phi_V}
\phi=\pm \frac{1}{\mathrm{V}}+\text{ const.}
\end{equation}
When static charged dust matter is included, the above conditions imply that
charge density must be equal to the mass density. An interesting feature of conformastat metrics
is that they permit the construction of complete asymmetrical relativistic configurations of matter in equibrium.
With metric coefficients given by (\ref{eq_choice}), we find the following expressions for the energy density 
and pressures Eq.\ (\ref{eq_disk_surf})
\begin{align}
\sigma &=-\frac{\mathrm{V}_{,z}}{2\pi\mathrm{V}^2} \mbox{,} \label{eq_sigma_con1} \\
P &=0 \mbox{.}
\end{align}
With condition (\ref{eq_phi_V}), Eq.\ (\ref{eq_disk_ch}) gives
\begin{equation}
\rho=-\frac{\mathrm{V}_{,z}}{2\pi\mathrm{V}^2} \mbox{,}
\end{equation}
where we have chosen the negative sign in Eq.\ (\ref{eq_phi_V}) that corresponds to
positive charges. Thus we have a distribution of charged dust matter where the charge
density is equal to the mass density on the plane $z=0$.
We use now a particular form of the function $V$ to construct disks and non-axisymmetric distributions of ECD.
\section{Aligned extreme Reissner-Nordstr\"{o}m black holes on the $z$ axis} \label{sec_aligned1}

\begin{figure}
\centering
\includegraphics[scale=0.8]{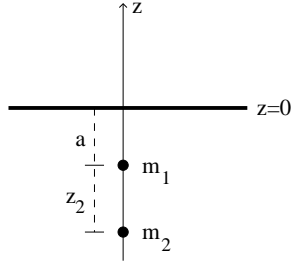}
\caption{Schematic drawing of the configuration used to construct charged dust disks.} \label{fig1}
\end{figure}

\begin{figure}
\centering
\includegraphics[scale=0.7]{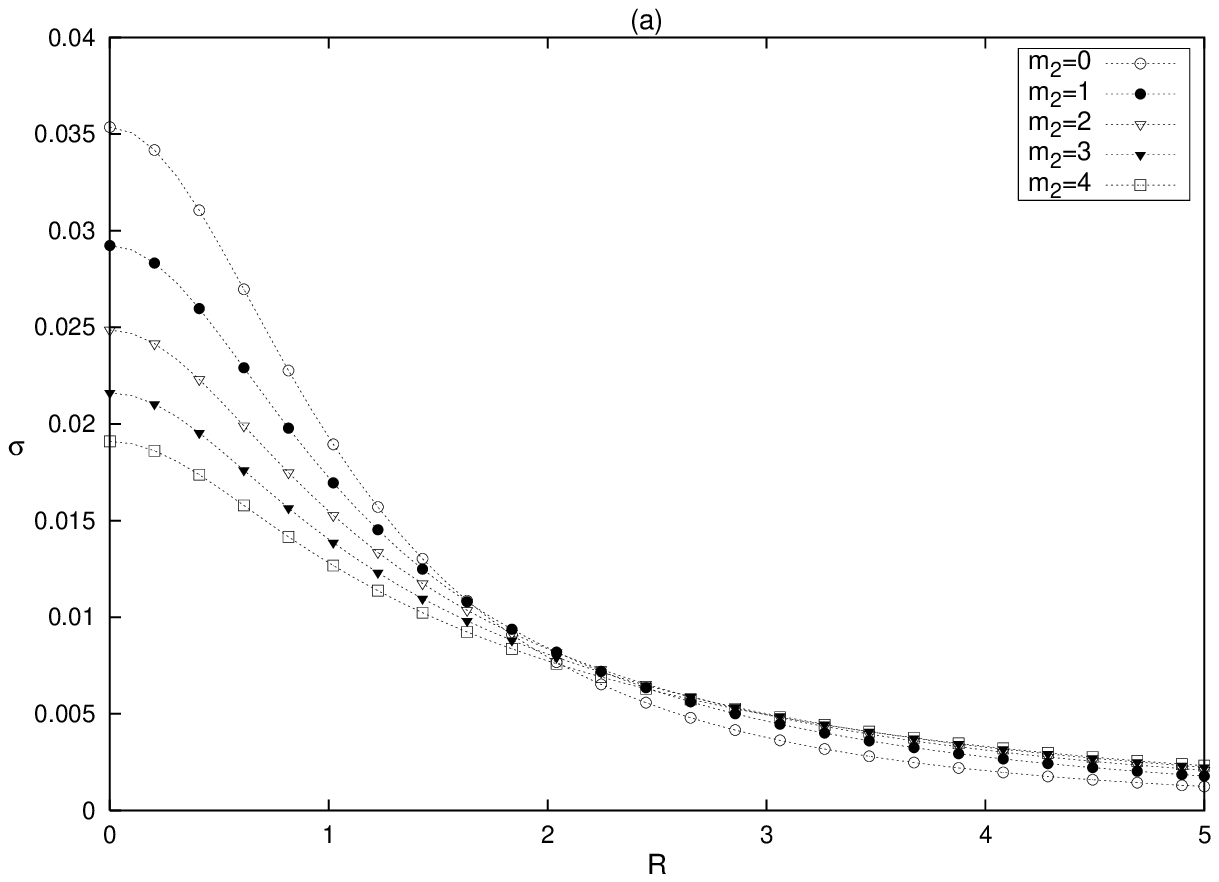}\\
\vspace{0.2cm}
\includegraphics[scale=0.7]{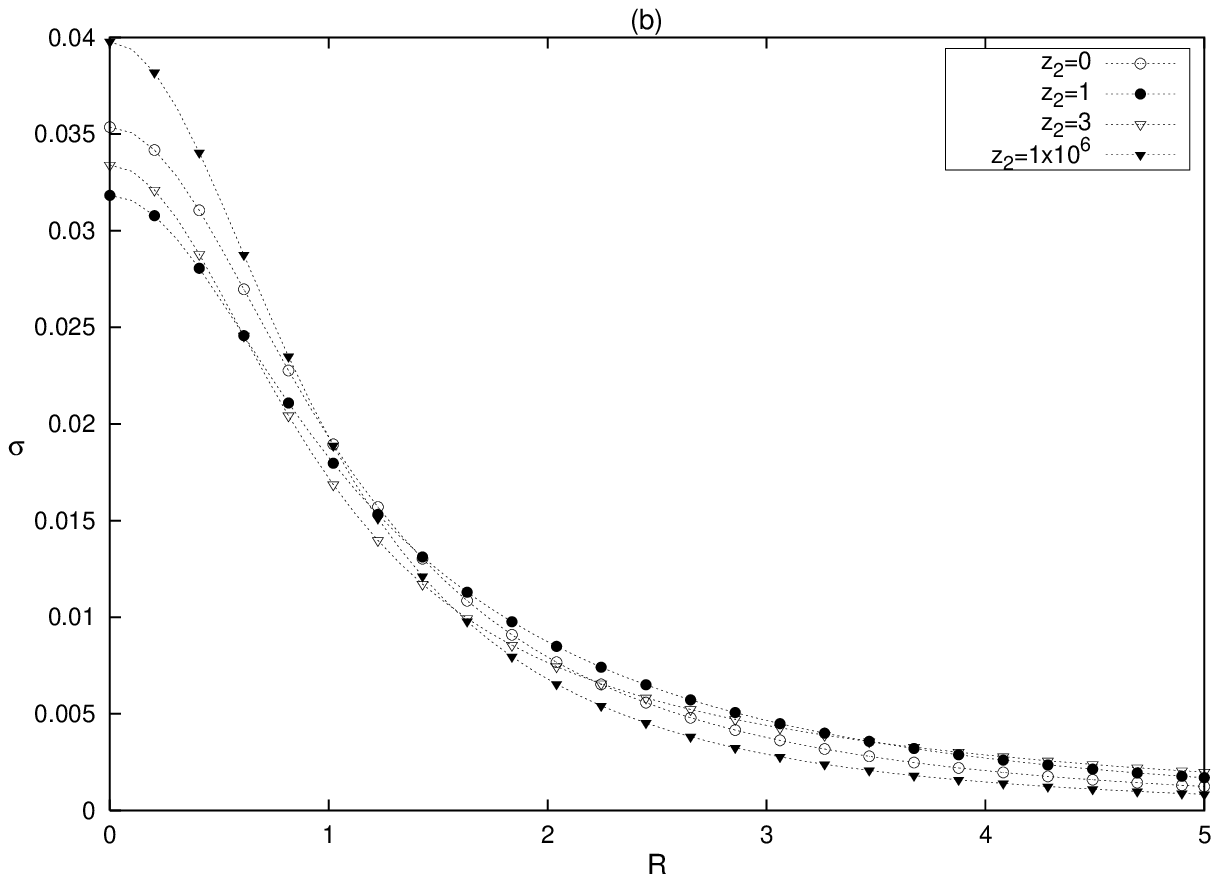}
\caption{Energy density $\sigma$ Eq.\ (\ref{eq_sigma_con2}) as a function of $R$. (a) $m_1=2$, $a=1$,
$z_2=1$ are fixed and $m_2$ is changed from $m_2=0$ to $m_2=4$. (b) $m_1=1$, $m_2=1$, $a=1$
are fixed and $z_2$ is changed: $z_2=0$, $1$, $3$ and $1\times 10^6$.} \label{fig2}
\end{figure}

Electrovacuum solutions representing extreme Reissner-Nordstr\"{o}m black holes in arbitrary
positions in static equilibrium have been found by Majumdar \cite{Majumdar} and Papapetrou \cite{Papapetrou}.  
Such solutions can be constructed because, as previously stated, the functions $V(x,y,z)$ must satisfy Laplace
equation and thus can be superposed.
Suppose we have two extreme Reissner-Nordstr\"{o}m black holes on $z$ axis, the first with mass $m_1$ at
$z=0$ and the second with mass $m_2$ at $z=-z_2$ (Fig.\ \ref{fig1}). For convenience we use cylindrical
coordinates $(t,R,z,\varphi)$. The function V representing the superposition is given by
\begin{equation} \label{eq_sup1}
\mathrm{V}=1+\frac{m_1}{\sqrt{R^2+z^2}}+\frac{m_2}{\sqrt{R^2+(z+z_2)^2}} \mbox{.}
\end{equation}
If we apply the ``displace, cut and reflect'' method to the system represented by Eq.\ (\ref{eq_sup1}), we
have for the energy density Eq.\ (\ref{eq_sigma_con1}):
\begin{equation} \label{eq_sigma_con2}
\sigma=\frac{m_1a\mathcal{R}_2^{3/2}+m_2(a+z_2)\mathcal{R}_1^{3/2}}{2\pi\sqrt{\mathcal{R}_1
\mathcal{R}_2}\left( \sqrt{\mathcal{R}_1\mathcal{R}_2}+m_1\sqrt{\mathcal{R}_2}+
m_2\sqrt{\mathcal{R}_1} \right)^2} \mbox{,}
\end{equation}
where $\mathcal{R}_1=R^2+a^2$ and $\mathcal{R}_2=R^2+(z_2+a)^2$. Note that $\sigma$ is always 
non-negative.
In Fig.\ \ref{fig2} we graph the energy density of the disk Eq.\ (\ref{eq_sigma_con2}) as a function of $R$ in
two situations: in (a) we fix $m_1=2$, $a=1$, $z_2=1$ and change $m_2$, in (b) we fix $m_1=1$, $m_2=1$,
$a=1$ and change $z_2$. In situation (a) we note that as $m_2$ is increased, the energy density is more
uniformly distributed along the radius. In (b) $z_2=0$ corresponds to one black hole with mass $m_1=2$
(curve $m_2=0$ in (a)). The energy density on $R=0$ assumes a minimum value when $z_2=1$, and 
then increases as $z_2$ is increased. The curve with $z_2=1\times 10^6$ (mass $m_2$ is very far away) 
corresponds to a disk generated solely by $m_1$.
\section{Aligned extreme Reissner-Nordstr\"{o}m black holes on the $y$ axis} \label{sec_aligned2}

\begin{figure}
\centering
\includegraphics[scale=0.8]{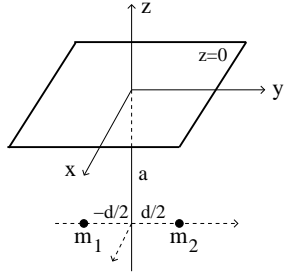}
\caption{Schematic drawing of the configuration used to construct asymmetric charged dust 
distributions.} \label{fig3}
\end{figure}

\begin{figure}
\centering
\includegraphics[scale=0.7]{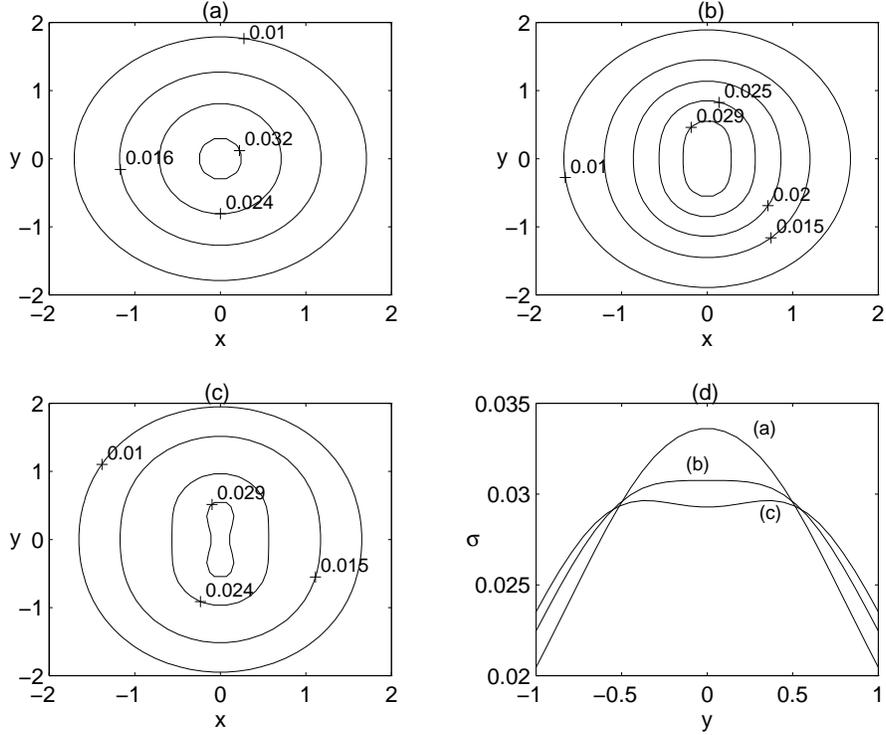}
\caption{Level curves of the energy density Eq.\ (\ref{eq_sigma_con3}), with $m_1=1$, $m_2=1$, $a=1$ and
(a) $d=0.5$, (b) $d=0.85$, (c) $d=1$. (d) The energy density is plotted along $x=0$, 
showing the extreme points $y_c$ for cases (a)--(c).} \label{fig4}
\end{figure}

Now we take two extreme Reissner-Nordstr\"{o}m black holes, with masses $m_1$ and $m_2$, located
at $(x,y,z)$ coordinates $(0,d/2,0)$ and $(0,-d/2,0)$, respectively (Fig.\ \ref{fig3}). The function V is
written as
\begin{equation} \label{eq_sup2}
\mathrm{V}=1+\frac{m_1}{\sqrt{x^2+(y-\frac{d}{2})^2+z^2}}+\frac{m_2}{\sqrt{x^2+(y+\frac{d}{2})^2
+z^2}} \mbox{.}
\end{equation}
Using again the ``displace, cut and reflect'' method, we obtain an expression for
the energy density Eq.\ (\ref{eq_sigma_con1}):
\begin{equation} \label{eq_sigma_con3}
\sigma=\frac{a}{2\pi}\frac{m_1\mathsf{R}_2^{3/2}+m_2\mathsf{R}_1^{3/2}}{\sqrt{\mathsf{R}_1
\mathsf{R}_2}\left( \sqrt{\mathsf{R}_1\mathsf{R}_2}+m_1\sqrt{\mathsf{R}_2}+
m_2\sqrt{\mathsf{R}_1} \right)^2} \mbox{,}
\end{equation}
where $\mathsf{R}_1=x^2+(y-d/2)^2+a^2$ and $\mathsf{R}_2=x^2+(y+d/2)^2+a^2$. Eq.\ (\ref{eq_sigma_con3})
represents a non-axisymmetric distribution of matter on the $z=0$ plane. In general it is symmetric only
with respect to the $y$ axis, but when $m_1=m_2$ it is also symmetric with respect to the $x$ axis. The extreme points $(x_c,y_c)$ of
the energy density, given by $\vec{\nabla}\sigma=0$, are $x_c=0$ and the roots of:
\begin{align}
2& \left( \frac{m_1}{\mathsf{R}_{1c}^{3/2}}+\frac{m_2}{\mathsf{R}_{2c}^{3/2}} \right)
\left[ \frac{m_1(y_c-\frac{d}{2})}{\mathsf{R}_{1c}^{3/2}}+\frac{m_2(y_c+\frac{d}{2})}{\mathsf{R}_{2c}^{3/2}} \right]
-3\left[ \frac{m_1(y_c-\frac{d}{2})}{\mathsf{R}_{1c}^{5/2}} \right. \notag \\
& \left. + \frac{m_2(y_c+\frac{d}{2})}{\mathsf{R}_{2c}^{5/2}} \right]
\left( 1+\frac{m_1}{\sqrt{\mathsf{R}_{1c}}}+\frac{m_2}{\sqrt{\mathsf{R}_{2c}}}\right)=0 \mbox{,} \label{eq_extr1}
\end{align}
where $\mathsf{R}_{1c}=(y_c-d/2)^2+a^2$ and $\mathsf{R}_{2c}=(y_c+d/2)^2+a^2$. For $m_1=m_2$ we note that 
$y_c=0$ is always a root of Eq.\ (\ref{eq_extr1}). Fig.\ \ref{fig4}(a)--(c)
is a contour plot of Eq.\ (\ref{eq_sigma_con3}) for $m_1=1$, $m_2=1$, $a=1$ and (a) $d=0.5$,
(b) $d=0.85$, (c) $d=1$. We note that as $d$ is increased, $y_c=0$ changes from a maximum to a local minimum
point, and two other symmetrical extreme points appear. This
is seen in Fig.\ \ref{fig4}(d), where $\sigma$ is plotted along $x=0$ for cases (a), (b) and (c). The transition of the
extreme point $y_c=0$ from a local maximum to a local minimum is determined by $\sigma_{,yy}=0$ evaluated
at $(x_c,y_c)=(0,0)$. In the particular case when $m_1=m_2=a=k$, this becomes
\begin{equation} \label{eq_trans}
12k^4-9d^2k^2-3d^4+4k(k^2-2d^2)\sqrt{4k^2+d^2}=0 \mbox{.}
\end{equation}
Eq.\ (\ref{eq_trans}) has a real root
\begin{equation}
d=k\frac{\sqrt{2+2\sqrt{145}}}{6} \approx 0.8511k \mbox{.}
\end{equation}

\begin{figure}
\centering
\includegraphics[scale=0.7]{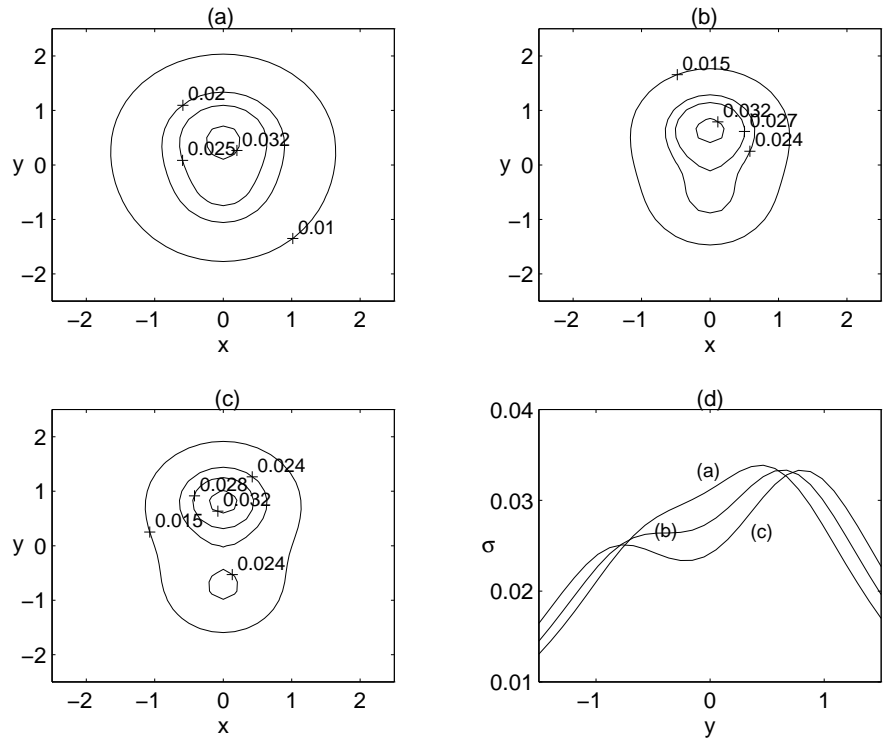}
\caption{Level curves of the energy density Eq.\ (\ref{eq_sigma_con3}), with $m_1=1$, $m_2=0.5$, $a=1$ and
(a) $d=1$, (b) $d=1.3$, (c) $d=1.6$. (d) The energy density is plotted along
$x=0$, showing the extreme points $y_c$ for cases (a)--(c).} \label{fig5}
\end{figure}

Fig.\ \ref{fig5}(a)--(c) is another contour plot of the energy density with $m_1=1$, $m_2=0.5$,
$a=1$ and (a) $d=1$, (b) $d=1.3$ and (c) $d=1.6$. For $d\lessapprox 1.3$ there is one maximum extreme point
$y_c$. When $d\gtrapprox 1.3$, there appear two maximum and one minimum extreme points $y_c$, all in
asymmetrical positions. This is pictured in Fig.\ \ref{fig5}(d).

\begin{figure}
\centering
\includegraphics[scale=0.7]{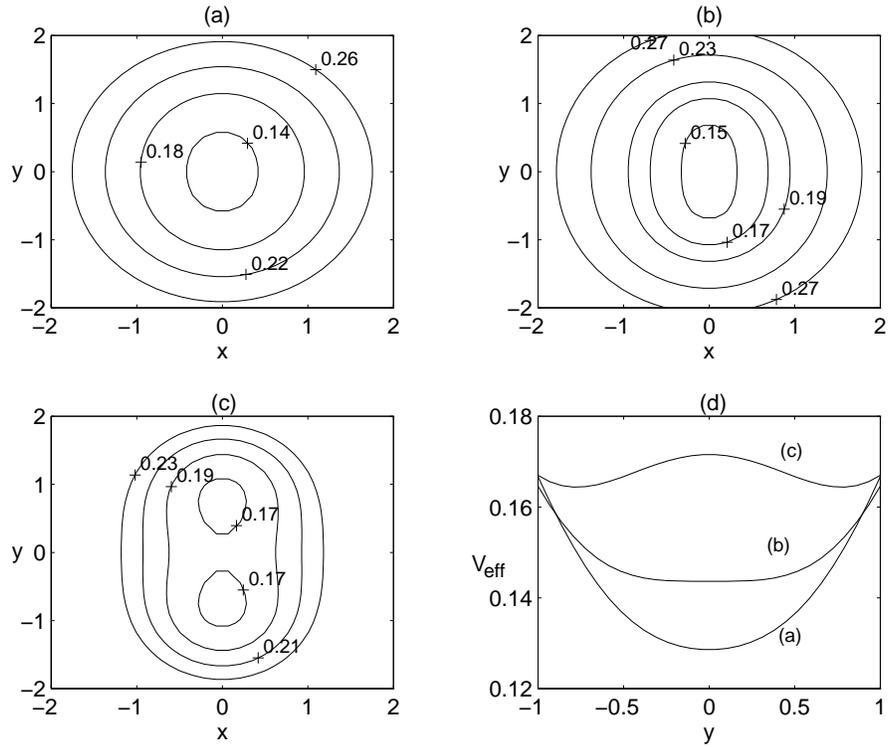}
\caption{Level curves of the effective potential Eq.\ (\ref{eq_pot_eff2}), with $m_1=1$, $m_2=1$, $a=1$ and
(a) $d=1$, (b) $d=1.4$, (c) $d=2$. (d) The effective potential is plotted along
$x=0$, showing the extreme points $y_c$ for cases (a)--(c).} \label{fig6}
\end{figure}

It is also interesting to study the effective potential of geodesic motion of neutral test particles for
this kind of matter distribution. The Lagrangean associated with the metric (\ref{eq_metric1}) is
\begin{equation} \label{eq_lagran}
2\mathcal{L}=\mathrm{V}^{-2}\dot{t}^2-\mathrm{V}^2(\dot{x}^2+\dot{y}^2+\dot{z}^2) \mbox{,}
\end{equation}
where the dots indicate derivatives with respect to the parameter $s$. One of Lagrange's equations 
gives
\begin{equation} \label{eq_energy}
\frac{d}{ds}\left( \frac{\partial \mathcal{L}}{\partial \dot{t}}\right) -\frac{\partial \mathcal{L}}{\partial t}=0
\Rightarrow V^{-2}\dot{t}=E=\text{const.}
\end{equation}
For time-like geodesic motion on the $xy$ plane, we may write
\begin{equation} \label{eq_en_cons}
1=\mathrm{V}^{-2}\dot{t}^2-\mathrm{V}^2(\dot{x}^2+\dot{y}^2) \rightarrow
\dot{x}^2+\dot{y}^2+\frac{1}{V^2}=E^2 \mbox{.}
\end{equation}
Thus we can define an effective potential $V_{\mathrm{eff.}}$ as
\begin{equation} \label{eq_pot_eff1}
V_{\mathrm{eff.}}=\frac{1}{\mathrm{V}^2} \mbox{.}
\end{equation}
For our matter distribution, Eq.\ (\ref{eq_pot_eff1}) reads
\begin{equation} \label{eq_pot_eff2}
V_{\mathrm{eff.}}=\frac{1}{\left[ 1+\frac{m_1}{\sqrt{x^2+(y-\frac{d}{2})^2+a^2}}+
\frac{m_2}{\sqrt{x^2+(y+\frac{d}{2})^2+a^2}} \right]^2} \mbox{.}
\end{equation}
Calculating $\vec{\nabla}V_{\mathrm{eff.}}=0$, one obtains the extreme points $x_c=0$ and the roots of
\begin{equation}
\frac{m_1(y_c-\frac{d}{2})}{\left[\left( y_c-\frac{d}{2}\right)^2+a^2 \right]^{3/2}}+
 \frac{m_2(y_c+\frac{d}{2})}{\left[\left( y_c+\frac{d}{2}\right)^2+a^2 \right]^{3/2}}=0 \mbox{.}
\label{eq_extr2}
\end{equation}
For $m_1=m_2$, $y_c=0$ is always a root of Eq.\ (\ref{eq_extr2}). Fig.\ \ref{fig6}(a)--(c)
is a contour plot of Eq.\ (\ref{eq_pot_eff2}) for $m_1=1$, $m_2=1$, $a=1$ and (a) $d=1$,
(b) $d=1.4$, (c) $d=2$. As $d$ is increased, $y_c=0$ changes from a maximum to a local minimum point, and 
two other symmetrical extreme points appear. This
is seen in Fig.\ \ref{fig6}(d), where $V_{\mathrm{eff.}}$ is plotted along $x=0$ for cases (a), (b) and (c). The transition of the
extreme point $y_c=0$ from a local maximum to a local minimum is determined by $(V_{\mathrm{eff.}})_{,yy}=0$ evaluated
at $(x_c,y_c)=(0,0)$. Setting $m_1=m_2=m$, we obtain
\begin{equation}
(2a^2-d^2)(4m+\sqrt{d^2+4a^2})=0 \mbox{,}
\end{equation}
which has a real root $d=a\sqrt{2}$.

\begin{figure}
\centering
\includegraphics[scale=0.7]{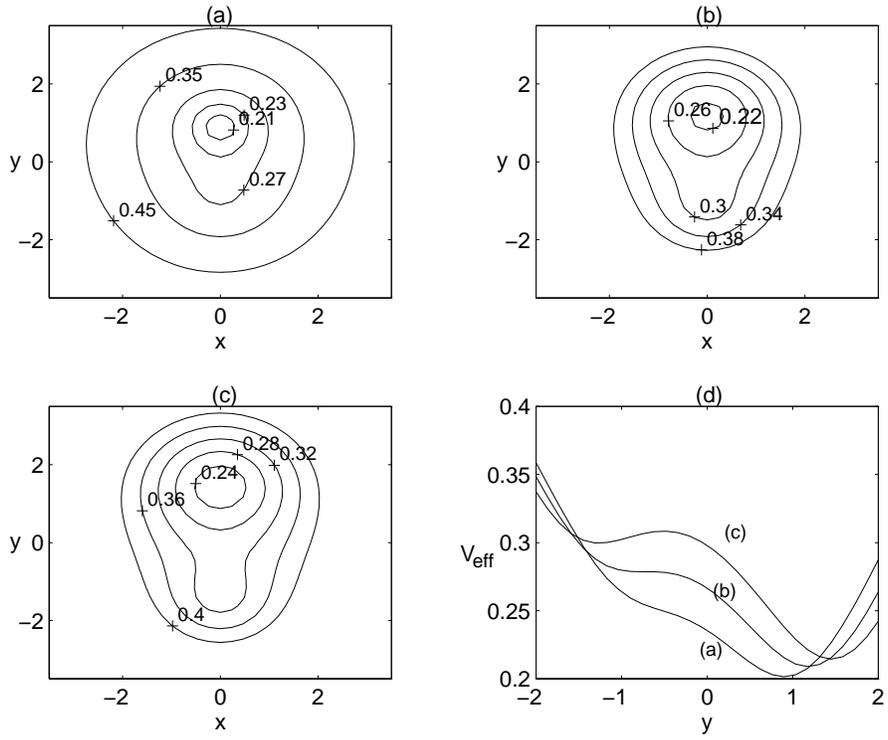}
\caption{Level curves of the effective potential Eq.\ (\ref{eq_pot_eff2}), with $m_1=1$, $m_2=0.5$, $a=1$ and
(a) $d=2$, (b) $d=2.5$, (c) $d=3$. (d) The effective potential is plotted along
$x=0$, showing the extreme points $y_c$ for cases (a)--(c).} \label{fig7}
\end{figure}

Fig.\ \ref{fig7}(a)--(c) is another contour plot of $V_{\mathrm{eff.}}$ with $m_1=1$, $m_2=0.5$,
$a=1$ and (a) $d=2$, (b) $d=2.5$ and (c) $d=3$. For $d\lessapprox 2.5$ there is one minimum extreme point
$y_c$. When $d\gtrapprox 2.5$, there appear two minimum and one maximum extreme points $y_c$, all in
asymmetrical positions. This is pictured in Fig.\ \ref{fig7}(d).
\section{Discussion} \label{sec_disc}

We applied the ``displace, cut and reflect'' method to a conformastat form of the metric to
generate disks made of electrically counterpoised dust (ECD) in equilibrium. The seed metric used was two
extreme Reissner-Nordstr\"{o}m black holes placed along the symmetry axis. Non-axisymmetric matter
distribution of ECD on the $z=0$ plane were obtained from the metric of two extreme Reissner-Nordstr\"{o}m
black holes located along the $y$ axis. To our knowledge no such configurations of matter
satisfying Einstein's field equations exist in the literature. We studied the variation of the energy density distribution by
varying the coordinate distance between the black holes as well as their masses. For equal black 
hole masses the energy density distribution is symmetric with respect to both the $x$ and $y$ axes, while for unequal
masses it possesses symmetry only with respect to the $y$ axis. The effective
potential for geodesic motion of neutral test particles on the plane of such a matter distribution
was also analysed. 

Of course one could extend the procedure used in this paper to generate complete
asymmetric planar configurations of ECD by putting $N>2$ extreme Reissner-Nordstr\"{o}m black holes in arbitrary
positions on the $xy$ plane and then applying the ``displace, cut and reflect'' method.

\bigskip
\centerline{\large{\textbf{Acknowledgments}}}
\bigskip

D.\ V.\ thanks CAPES for financial support. P.\ S.\ L.\ thanks FAPESP and CNPq for financial support.

\end{document}